\documentclass[aps,prb,showpacs,twocolumn,superscriptaddress]{revtex4-1}

\usepackage{graphicx} 
\usepackage{color}
\usepackage{array}
\usepackage{mathrsfs}

\usepackage{ulem} 

\begin{document}

\title{Static and Dynamic Magnetic Properties of spin-$\frac{1}{2}$ Inequilateral Diamond-Chain Compounds $A_3$Cu$_3$AlO$_2$(SO$_4$)$_4$ ($A$=K, Rb, and Cs)}



\author{Katsuhiro Morita}
\email[e-mail:]{katsuhiro.morita@rs.tus.ac.jp}
\affiliation{Department of Applied Physics, Tokyo University of Science, Tokyo 125-8585, Japan}

\author{Masayoshi Fujihala}
\affiliation{Department of Physics, Tokyo University of Science, Shinjuku, Tokyo 162-8601, Japan}

\author{Hiroko Koorikawa}
\affiliation{Department of Physics, Tokyo University of Science, Shinjuku, Tokyo 162-8601, Japan}

\author{Takanori Sugimoto}
\affiliation{Department of Applied Physics, Tokyo University of Science, Tokyo 125-8585, Japan}

\author{Shigetoshi Sota}
\affiliation{RIKEN Advanced Institute for Computational Science (AICS), Kobe, Hyogo 650-0047, Japan}

\author{Setsuo Mitsuda}
\affiliation{Department of Physics, Tokyo University of Science, Shinjuku, Tokyo 162-8601, Japan}

\author{Takami Tohyama}
\affiliation{Department of Applied Physics, Tokyo University of Science, Tokyo 125-8585, Japan}


\date{\today}

\begin{abstract}
Spin-$\frac{1}{2}$ compounds $A_3$Cu$_3$AlO$_2$(SO$_4$)$_4$ ($A$=K, Rb, and Cs) have one-dimensional (1D) inequilateral diamond-chains. We analyze the temperature dependence of the magnetic susceptibility and determine the magnetic exchange interactions. In contrast to azurite, a dimer is formed on one of the sides of the diamond. From numerical analyses of the proposed model, we find that the dimer together with a nearly isolated 1D Heisenberg chain characterize magnetic properties including magnetization curve and magnetic excitations. This implies that a dimer-monomer composite chain without frustration is a good starting point for describing these compounds.
\end{abstract}


\maketitle
\section{Introduction}
Highly frustrated quantum magnets provide various exotic ground states such as gapless spin-liquid and gapped singlet dimer phases~\cite{Balents2010,Lacroix2011,Imai2016}. In a magnetic field, the magnets exhibit magnetization plateaus because of the competition of frustration and quantum fluctuations. The typical constituent of frustrated magnets is a triangular unit of spin with antiferromagnetic (AFM) interaction for each bond. The spin-$\frac{1}{2}$ diamond-chain where the triangular unit is connected linearly thus is regarded as a typical highly frustrated system in one dimension~\cite{Takano1996,Okamoto1999,Okamoto2003}.

azurite Cu$_3$(CO$_3$)$_2$(OH)$_2$ has originally been suggested to be a spin-$\frac{1}{2}$ distorted diamond-chain with AFM interactions for three bonds of a triangular unit~\cite{Kikuchi2005}. A recent theoretical approach based on density functional theory together with numerical many-body calculations has proposed a microscopic model of azurite with less frustrated interactions~\cite{Jeschke2011}: Two of three Cu$^{2+}$ spins are coupled strongly by AFM interaction $J_2$ [see Fig.~\ref{structure}(b)] to form a dimer singlet, whereas another spin consists of a monomer spin that is weakly connected to neighboring monomer spins by AFM interaction $J_\mathrm{m}$, which has been indicated in the early stage of research in azurite~\cite{Kikuchi2005}. This model, including the two energy scales of $J_2$ and $J_\mathrm{m}$, has nicely reproduced the double-peak structures observed in the magnetic susceptibility (a peak at 5~K and a broad peak at 23~K)~\cite{Kikuchi2005} and the specific heat~\cite{Kikuchi2005,Rule2008}. In a magnetic field, the $1/3$ magnetization plateau~\cite{Kikuchi2005} is interpreted as a result of almost fully polarized monomer spins and bounded dimer spins~\cite{Jeschke2011}. The model predicts a gapless low-energy spin excitation originating from a spin-liquid behavior due to an effective spin-$\frac{1}{2}$ Heisenberg chain~\cite{Honecker2011}. However, three-dimensional magnetic interactions in azurite cause a magnetic order below 1.85~K. 

Recently, a new highly one-dimensional (1D) diamond-chain compound K$_3$Cu$_3$AlO$_2$(SO$_4$)$_4$ has been reported~\cite{Fujihala2015}. In this compound, the magnetic susceptibility exhibits a double-peak structure similar to azurite, but the temperatures of the peaks (50 and 200~K) are one order of magnitude higher than those in azurite. Despite  such high characteristic temperatures, there is no magnetic order down to 0.5~K, indicating a possible spin-liquid ground state~\cite{Fujihala2015}. It is, thus, important to clarify common features characterizing the distorted diamond-chain compounds in both azurite and the new compound.

In this paper, we analyze the temperature dependence of the magnetic susceptibility in K$_3$Cu$_3$AlO$_2$(SO$_4$)$_4$ as well as newly synthesized compounds where Rb and Cs are substituted for K by using the finite-temperature Lanczos (FTL) method~\cite{Jaklic2000} and the exact diagonalization (ED) method. The estimated magnetic exchange interactions are found to form strong dimer bonds and monomer-monomer chains. This is similar to azurite, although the dimer-bond positions as well as their energy scales are different. The frustration is less effective in K$_3$Cu$_3$AlO$_2$(SO$_4$)$_4$ than in azurite, and the spin-liquid behavior at low temperatures is attributed to an effective spin-$\frac{1}{2}$ Heisenberg chain. Therefore, it is reasonable to conclude that diamond-chain compounds consisting of Cu$^{2+}$ are less frustrated materials and thus a good starting point for the compounds is a dimer-monomer composite structure. Based on the estimated exchange interactions in K$_3$Cu$_3$AlO$_2$(SO$_4$)$_4$, we predict the magnetization curve with the 1/3 plateau and inelastic neutron-scattering spectrum by density matrix renormalization group (DMRG) calculations. 

This paper is organized as follows. We describe the crystal structure of $A_3$Cu$_3$AlO$_2$(SO$_4$)$_4$ ($A=$~K, Rb, and Cs) and discuss this effective model in Sec.~\ref{sec:2}.
 In Sec.~\ref{sec:3}, we analyze the temperature dependence of the magnetic susceptibility of $A_3$Cu$_3$AlO$_2$(SO$_4$)$_4$ and determine the magnetic exchange interactions. The magnetization curve and dynamical spin structure factor in K$_3$Cu$_3$AlO$_2$(SO$_4$)$_4$ are shown in Sec.~\ref{sec:4}.
 Finally, a summary is given in Sec.~\ref{sec:5}.
\section{crystal structure and model}
\label{sec:2}
\begin{figure}[tb]
\includegraphics[width=60mm]{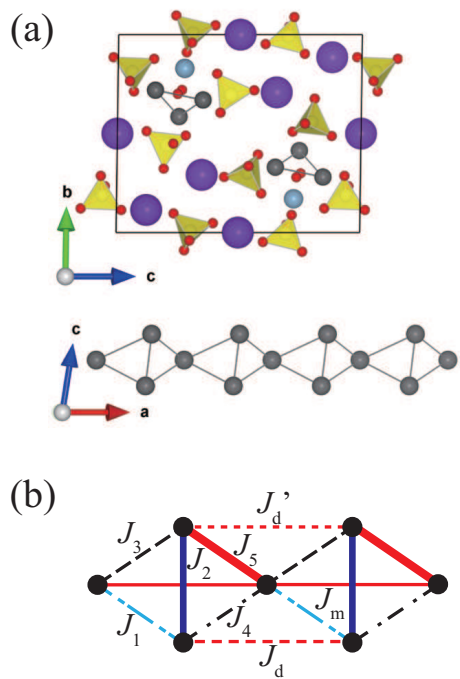}
\caption{(Color online) (a) Crystal structure of $A_3$Cu$_3$AlO$_2$(SO$_4$)$_4$ ($A=$~K, Rb, and Cs). The gray, purple, light blue, and red circles denote Cu, $A$, Al, and O atoms, respectively. The tetrahedrons with the red dots at the corners denote (SO$_4$). The inequilateral diamond-chains run along the $a$ axis. (b) Effective spin model of $A_3$Cu$_3$AlO$_2$(SO$_4$)$_4$. The circles represent Cu$^{2+}$ ions with spin 1/2. The blue broken, dark blue solid, black broken, black dashed-dotted, red thick solid, red thin solid, red dashed, and red dotted lines denote the exchange interactions $J_1$, $J_2$, $J_3$, $J_4$, $J_5$, $J_\mathrm{m}$, $J_\mathrm{d}$, and $J_\mathrm{d}'$, respectively. 
\label{structure}}
\end{figure}

The crystal structure of $A_3$Cu$_3$AlO$_2$(SO$_4$)$_4$
is shown in Fig.~\ref{structure}(a). The diamond-chains composed of Cu$^{2+}$ ions are formed along the $a$ axis. Since the diamonds are inequilateral as discussed below, exchange interactions for the nearest-neighbor bonds [$J_1$ to $J_5$ as shown in Fig.~\ref{structure}(b)] are not necessarily the same. In addition, we consider exchange interactions connecting neighboring triangular units, denoted by $J_\mathrm{m}$, $J_\mathrm{d}$, and $J'_\mathrm{d}$ in Fig.~\ref{structure}(b). We note that only $J_\mathrm{m}$ is taken into account in azurite. In the present compounds there are possible paths for the $J_\mathrm{d}$ and $J'_\mathrm{d}$ bonds through SO$_4$ units. Since the surrounding components of the three $J_\mathrm{m}$, $J_\mathrm{d}$, and $J'_\mathrm{d}$ bonds are similar to each other, we assume that $J_\mathrm{m}=J_\mathrm{d}=J'_\mathrm{d}$. 

The effective spin Hamiltonian for $A_3$Cu$_3$AlO$_2$(SO$_4$)$_4$ under the external magnetic-field $H$ is thus given by
\begin{eqnarray} 
\mathscr{H} &=& \sum_{\langle i,j \rangle } J_{ij}\mathbf{S}_i \cdot \mathbf{S}_j - g\mu_{\rm B}H\sum_i S^{z}_i,
\end{eqnarray}
where $\mathbf{S}_i$ is the spin-$\frac{1}{2}$ operator, $J_{ij}$ is the exchange interaction corresponding to the bonds shown in Fig.~\ref{structure}(b), $\mu_{\rm B}$ is the Bohr magneton, and $g$ is the gyromagnetic ratio.

Before fitting calculated magnetic susceptibilities to experimental ones, we need to roughly evaluate the value of exchange interactions. From the crystal structure analysis of K$_3$Cu$_3$AlO$_2$(SO$_4$)$_4$, the average Cu-O-Cu angle is estimated to be $104.7^\circ, 95.2^\circ, 102.5^\circ, 105.8^\circ$, and $132.0^\circ$ for the $J_1$, $J_2$, $J_3$, $J_4$, and $J_5$ bond, respectively~\cite{FujihalaUnpublished}. Since the Cu-O-Cu angle significantly influences on the value of the exchange interactions~\cite{Mizuno1998}, the variation of the angles can give strong bond-dependent exchange interactions. According to the angle-dependent exchange interaction of cuprates~\cite{Mizuno1998}, $J_5$ with the largest angle is expected to be an AFM interaction with very roughly $\sim$500~K, whereas $J_2$ with the smallest angle is to be ferromagnetic (FM) ($\sim -$100~K). The values of the exchange interactions for other bonds are expected to be in between $J_2$ and $J_5$. For simplicity, we take $J_1=J_3=J_4$ because of similar Cu-O-Cu angles. We emphasize that the side of the $J_1$ bond and the side of the $J_5$ bond, which are opposite sides of a diamond, are inequivalent. This means that the diamond is distorted making opposite sides inequivalent, i.e., an inequilateral diamond. We thus call $A_3$Cu$_3$AlO$_2$(SO$_4$)$_4$ the inequilateral diamond-chain compound.

\begin{figure}[tb]
\includegraphics[width=75mm]{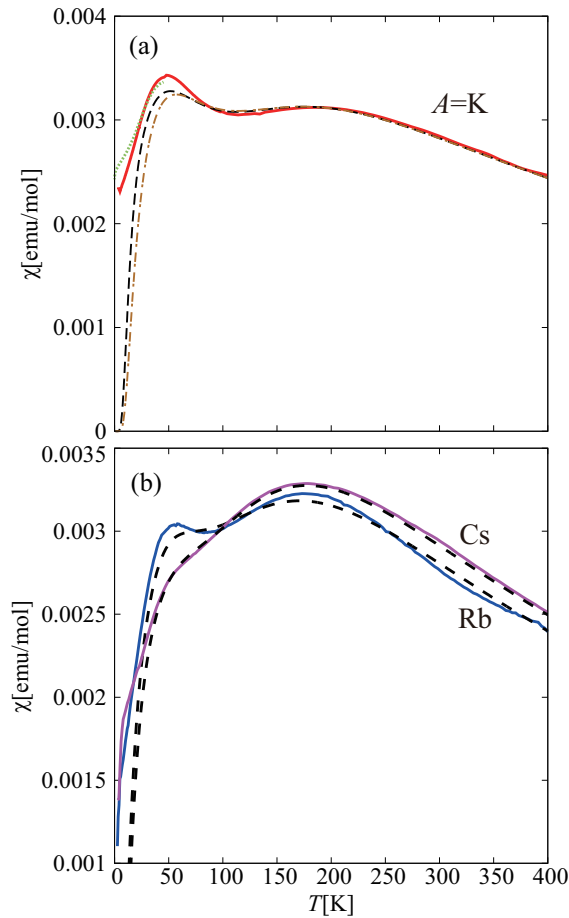}
\caption{(Color online) Temperature dependence of spin susceptibility in $A_3$Cu$_3$AlO$_2$(SO$_4$)$_4$ for (a) $A=$~K and (b) $A=$~Rb and Cs. The red solid lines show the experimental datad.
The black dashed lines represent the fitted data obtained by the FTL method for a 24-spin inequilateral diamond-chain. Note that the error of the FTL is within the width of the line. In (a), an ED result for an 18-spin chain is denoted by the brown dot-dashed line, and the Bethe-ansatz solution for the infinite Heisenberg chain is plotted by the green dotted line. The parameters are listed in Table~\ref{para}.
\label{chi}}
\end{figure} 
\section{magnetic susceptibilities}
\label{sec:3}
Taking into account this initial guess for the exchange interactions, we first investigate the temperature dependence of the spin susceptibility for K$_3$Cu$_3$AlO$_2$(SO$_4$)$_4$ by performing the FTL calculations for a 24-site diamond periodic chain [eight triangular units (the total number of site $N=8\times3$)] together with the ED calculations for an 18-site diamond periodic chain. The calculated spin susceptibility $\chi$ is compared to the experimental data~\cite{Fujihala2015} obtained by subtracting the diamagnetic susceptibility $\chi_\mathrm{dia}$, the impurity-spin paramagnetic susceptibility $\chi_\mathrm{imp}$, and the Van Vleck paramagnetic susceptibility $\chi_\mathrm{VV}$~\cite{VanVleck} from the experimentally observed magnetic susceptibility. Figure~\ref{chi}(a) shows the experimental result (red solid line) and fitted results (black dashed line for $N=24$ and brown dot-dashed line for $N=18$) of $\chi$ for K$_3$Cu$_3$AlO$_2$(SO$_4$)$_4$. The parameter values are listed in Table~\ref{para}. With increasing system size from $N=18$ to 24, the fitted results systematically approache the experimental one, indicating that the deviation from the experiment at $T<70$~K is due to the finite size effect. We find that the double-peak structure is reproduced clearly: A broad peak at 200~K comes from large $J_5$ forming a dimer on the corresponding bond, whereas the low-temperature peak at 50~K is attributed to a 1D Heisenberg interaction with positive $J_\mathrm{d}$ being similar to the case of azurite. The other parameters only affect the heights of the two peaks. Since low-temperature $\chi$ for $N=24$ below 30~K agrees with $\chi$ for an eight-site Heisenberg chain with the exchange interaction $J_\mathrm{d}$, it is naturally expected that the low-temperature $\chi$ in experiment is reproduced by the exact $\chi$ for the 1D Heisenberg model. In fact, the Bethe-ansatz solution of the Heisenberg model~\cite{Griffiths1964} shown as the dotted line in Fig.~\ref{chi}(a) well reproduces the experimental data, although some deviations probably due to the uncertainty of $\chi_\mathrm{imp}$ remain.

The obtained value of $J_5$ is the largest and 15 times larger than the maximum interaction in azurite ($\sim$33~K). Similarly $J_\mathrm{d}$ ($=J'_\mathrm{d}=J_\mathrm{m}$) in $A=$~K is 16 times larger than $J_\mathrm{m}$ in azurite ($\sim$4.62~K). Another important difference appears on the $J_2$ bond: $J_2$ is FM in K$_3$Cu$_3$AlO$_2$(SO$_4$)$_4$, whereas the dimer is located on the bond in azurite. It is also remarkable that there is only weak frustration in the diamond of K$_3$Cu$_3$AlO$_2$(SO$_4$)$_4$ since the magnitude of the FM $J_4$ interaction inducing frustration in a triangle is very small as compared with two other interactions $J_2$ and $J_5$.

\begin{table}[tb]
\caption{The exchange interactions and the gyromagnetic ratio obtained by fitting calculated $\chi$ to experimental ones.}
  \begin{tabular}{|c|c c c c|c|} \hline \hline
    A  &  $J_1=J_3=J_4$ & $J_2$ & $J_5$ & $J_\mathrm{m}=J_\mathrm{d}=J'_\mathrm{d}$
 & $g$\\ \hline 
    K & -30 & -300 & 510 & 75 & 2.14 \\ \hline
    Rb & -17 & -252 & 462 & 84 & 2.12  \\ \hline
    Cs & -19 & -238 & 456 & 95 & 2.17 \\ \hline \hline
  \end{tabular}
\label{para}
\end{table}

To confirm the magnetic interactions on $A$-site substituted compounds, we synthesized a single phase crystal with $A=$~Rb and Cs by a solid-state reaction in which high-purity $A_2$SO$_4$, CuO, CuSO$_4$ and AlK(SO$_4$)$_2$ powder were mixed with a molar ratio of 1 : 2 : 1 : 1. The mixture was heated at 600$^\circ$C for three days and then slowly cooled in air.

We fit calculated $\chi$ to the experimental ones for $A=$~Rb and Cs in Fig.~\ref{chi}(b). The two-peak structure is less pronounced but visible for $A=$~Cs. From the estimated parameter values of the exchange interactions listed in Table~\ref{para}, we find that $J_5$ for $A=$~Rb and Cs is 10\% smaller than that for $A=$~K. Actually the broad peak position shifts to a lower temperature by nearly the same amount. 
In contrast, $J_\mathrm{d}$ ($=J'_\mathrm{d}=J_\mathrm{m}$) increases from $A=$~K and Rb to Cs, inducing a slight shift of the low-temperature peak to a high-temperature peak. Other parameters with FM interactions reduce their magnitude from $A=$~K to Rb and Cs. These material-dependent changes in the interactions indicate a small change in Cu-O-Cu bond angles between K and Rb (Cs). A detailed crystal structure analysis will be necessary to confirm this and remains a future problem.

\section{magnetization curve and Dynamical spin structure factor}
\label{sec:4}
\begin{figure}[tb]
\includegraphics[width=80mm]{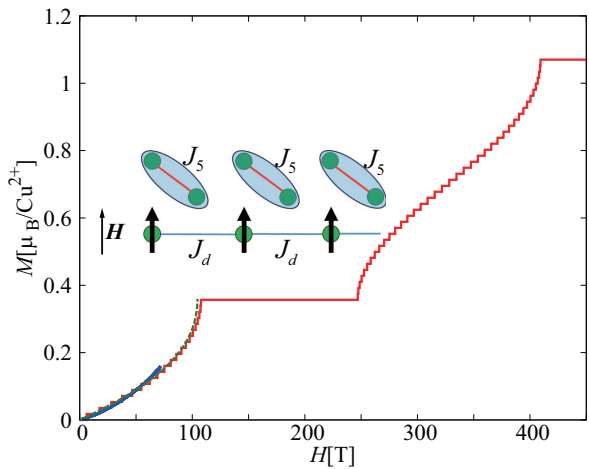}
\caption{(Color online) Magnetization curve for K$_3$Cu$_3$AlO$_2$(SO$_4$)$_4$. The red jagged solid line is a calculated curve by DMRG at zero temperature for a 120-site periodic chain with the exchange interactions estimated from $\chi$. The blue solid line represents the experimental result under the magnetic field up to 72~T at 4.2~K~\cite{Fujihala2015}. 
The green dashed line is the exact magnetization curve for the 1D Heisenberg model.
The inset is a schematic of the spin configuration at the 1/3 plateau with the dimers formed by $J_5$ and the 1D chain with $J_\mathrm{d}$ whose spins are ferromagnetically aligned with the direction of the applied magnetic field $H$.
\label{M-H}}
\end{figure}

To confirm the validity of the estimated exchange interactions, we calculate the magnetization curve for K$_3$Cu$_3$AlO$_2$(SO$_4$)$_4$ and compare it with available experimental data~\cite{Fujihala2015}. The magnetization curve is calculated by DMRG for a [$N=120(=40\times 3)$]-site periodic chain at zero temperature. The number of states kept in the DMRG calculation is $m=300$, and the resulting truncation error is less than $2\times10^{-6}$. Figure~\ref{M-H} shows the calculated magnetization curve (red solid curve) as well as the experimental data (blue solid line) for a low magnetic field up to $H$=72~T~\cite{Fujihala2015}. The agreement with the experimental data is quite good. The magnetization near zero field is proportional to $H$, which is characteristic behavior in the 1D Heisenberg model and consistent with the fact that the low-energy scale is controlled by 1D interaction $J_\mathrm{d}$ as evidenced by good agreement with the exact magnetization curve (green dashed line) for the 1D Heisenberg model~\cite{Griffiths1964}. The calculated curve exhibits a magnetization plateau at the magnetization $M=1/3$ as expected. The 1/3 plateau starts from 108~T, which can be accessible by a pulse magnet experiment. Such an experiment is desired to confirm our proposed model. The calculated onset field of the 1/3 magnetization plateau is 119 and 130~T for $A=$~Rb and Cs, respectively (not shown here). The slight increase in the onset field as compared with the $A=$~K case is attributed to the increase in $J_\mathrm{d}$.

We also examine the dynamical spin structure factor for K$_3$Cu$_3$AlO$_2$(SO$_4$)$_4$, defined by
\begin{equation} 
S(q,\omega)=-\frac{1}{\pi N}\mathrm{Im} \left<0\right| S_{-q}^z \frac{1}{\omega-\mathscr{H}+E_0+i\eta} S_q^z \left|0\right>,
\end{equation}
where $q$ is the momentum for the triangular unit cell, $\left|0\right>$ is the ground state with energy $E_0$, $\eta$ is a broadening factor, and $S_q^z=N^{-1/2}\sum_i e^{iqR_i} S_i^z$ with $R_i$ being the position of spin $i$ and $S_i^z$ being the $z$ component of \mbox{\boldmath $S$}$_i$.
$S(q,\omega)$ is calculated by using the dynamical DMRG~\cite{Sota2010} for a ($N$=240)-site periodic chain (80 triangular cells). The truncation number is $m=400$, and the truncation error is less than $7\times10^{-3}$. The value of $\eta$ is taken to be 0.65~meV.

\begin{figure}[tb]
\includegraphics[width=80mm]{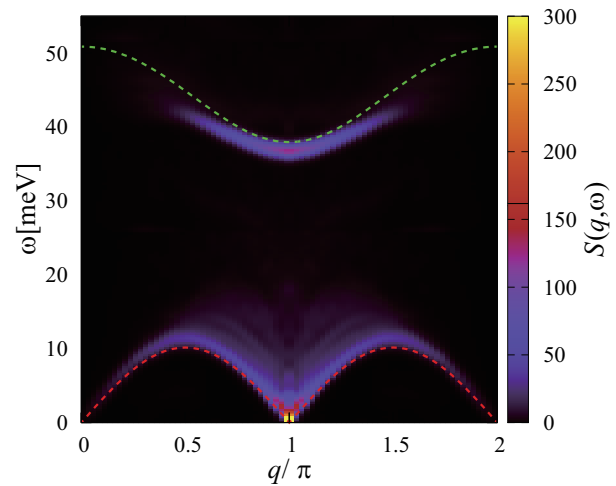}
\caption{(Color online) Dynamical spin structure factor $S(q,\omega)$ obtained by dynamical DMRG for a 240-site periodic chain with the exchange interactions for K$_3$Cu$_3$AlO$_2$(SO$_4$)$_4$.  The red dashed line represents $(\pi/2)J\left|\sin q\right|$ with $J=J_\mathrm{d}$, whereas the green dashed line represents $\omega_q=J_5 +J_\mathrm{m}\cos q+\frac{1}{4}J_\mathrm{m}^2/J_5(3-\cos 2q)$.
\label{DDMRG}}
\end{figure}

Figure~\ref{DDMRG} shows the contour plot of $S(q,\omega)$. At the low-energy region below 10~meV, we find a clear dispersive behavior fitted quite well by $(\pi/2)J\left|\sin q\right|$ with $J=J_\mathrm{d}$ (the red dashed line). This indicates that the lowest-energy branch comes from the 1D Heisenberg chain connected by the $J_\mathrm{d}$ bond. At the high-energy region around 40~meV, there is a dispersive structure having a minimum at $q=\pi$. This is nothing but the dispersion of a dimer predominantly formed on the $J_5$ bond. The dispersion relation is well reproduced by the second-order perturbation theory in terms of $J_\mathrm{m} (= J'_\mathrm{d})$ giving a dispersion of $\omega_q=J_5 +J_\mathrm{m}\cos q+\frac{1}{4}J_\mathrm{m}^2/J_5(3-\cos 2q)$ (the green dashed line)~\cite{Reigrotzki1994,Sushkov1998}, although there is a small deviation. Both the low-energy and high-energy structures should appear in inelastic neutron-scattering experiments. In fact, a preliminary experiment for the powder sample of K$_3$Cu$_3$AlO$_2$(SO$_4$)$_4$ has shown the corresponding structures~\cite{FujihalaUnpublished}.

\section{summary}
\label{sec:5}
We have examined the temperature dependence of the magnetic susceptibility for the inequilateral diamond-chain compound $A_3$Cu$_3$AlO$_2$(SO$_4$)$_4$ ($A=$~K, Rb, and Cs) both experimentally and theoretically. The systematic analyses for $A=$~K, Rb, and Cs clearly demonstrate that one of the bonds of the diamond has a strong AFM exchange interaction, producing a dimer. On the other hand, the bond shared by two triangles in the diamond is FM, in contrast to azurite where a dimer is formed on this bond. These behaviors are in accord with the angle dependence of the Cu-O-Cu bond. The dimer controls a high-temperature peak of the magnetic susceptibility as well as a high-energy dispersive structure in the dynamical spin structure factor. On the other hand, a low-energy peak in the magnetic susceptibility and low-energy excitations are controlled by monomers forming a 1D Heisenberg chain. Therefore, the dimer-monomer composite structure is a good starting point of diamond-type quantum spin compounds including azurites, in contrast to the original idea that the diamond-chain compounds are highly frustrated. Spin-liquid behaviors observed in the diamond-chain compounds thus are attributed to the presence of a 1D Heisenberg chain formed by the monomers. In $A_3$Cu$_3$AlO$_2$(SO$_4$)$_4$, the magnetization curve with the 1/3 plateau and inelastic neutron-scattering spectra separated by the two energy scales are expected as theoretically demonstrated. Experiments to confirm these predictions are in progress.

\begin{acknowledgments}
This work was supported, in part, by MEXT as a social and scientific priority issue (creation of new functional devices and high-performance materials to support next-generation industries (GCDMSI) to be tackled by using a post-K computer and by MEXT HPCI Strategic Programs for Innovative Research (SPIRE) (hp160222). The numerical calculation partly was carried out at the K Computer, Institute for Solid State Physics, The University of Tokyo and the Information Technology Center, The University of Tokyo. This work also was supported by Grants-in-Aid for Scientific Research (No. 26287079), Grants-in-Aids for Young Scientists (B) (No.~16K17753) from MEXT, Japan.
\end{acknowledgments}


\begin{thebibliography}{99}
\bibitem{VanVleck} $\chi_\mathrm{dia}$ is estimated from \cite{Bain2008}. $\chi_\mathrm{imp}$ is assumed to be Curie-Weiss-type, i.e., $\chi_\mathrm{imp} = n_\mathrm{imp}C/(T)$, where $C$ is the Curie constant for $S=1/2$ and we set $n_\mathrm{imp}=0.017$. The Van Vleck paramagnetic susceptibility was estimated using $\chi_\mathrm{VV}=(N\mu_\mathrm{B}^2/\lambda)\Delta g = 9.42 \times 10^{-4} \Delta g$ emu/mol, where $N$ is the number of Cu$^{2+}$ ions, $\Delta g=g-g_\mathrm{e}$ is the anisotropy of the gyromagnetic ratio, and $\lambda = 829$~cm$^{-1}$ is the spin-orbit coupling coefficient of Cu$^{2+}$.
\end{thebibliography}
\end{document}